\documentclass[twocolumn]{aastex63}
\usepackage{verbatim}
\usepackage{comment} 

\graphicspath{{./}{figures/}}
\received{January 25, 2023}
\submitjournal{ApJ}

\newcommand{\leff}{\lambda_{E}^{\rm eff}}
\definecolor{afcolor}{HTML}{b3443c}

\def\myr{{\rm Myr}}
\def\gyr{{\rm Gyr}} 

\def\NV{\hbox{N~$\scriptstyle\rm V $~}}

\def\gtsima{$\; \buildrel > \over \sim \;$}
\def\ltsima{$\; \buildrel < \over \sim \;$}
\def\gsim{\lower.5ex\hbox{\gtsima}} 
\def\lsim{\lower.5ex\hbox{\ltsima}}
\def\prosima{$\; \buildrel \propto \over \sim \;$} 
\def\simgt{\lower.5ex\hbox{\gtsima}} 
\def\simlt{\lower.5ex\hbox{\ltsima}} 
\def\simpr{\lower.5ex\hbox{\prosima}}
\def\msun{{\rm M}_{\odot}}
\def\zsun{{\rm Z}_{\odot}}
\def\lsun{{\rm L}_{\odot}}
\def\dsun{\mathcal{D}_{\odot}}

\begin{document}



\def\be{\begin{equation}}
\def\ee{\end{equation}}
\newcommand\code[1]{\textsc{\MakeLowercase{#1}}}
\newcommand\quotesingle[1]{`{#1}'}
\newcommand\quotes[1]{``{#1}"}
\def\gsim{\lower.5ex\hbox{\gtsima}} 
\def\lsim{\lower.5ex\hbox{\ltsima}} 
\def\gtsima{$\; \buildrel > \over \sim \;$} 
\def\ltsima{$\; \buildrel < \over \sim \;$} \def\gsim{\lower.5ex\hbox{\gtsima}} 
\def\lsim{\lower.5ex\hbox{\ltsima}} 
\def\simgt{\lower.5ex\hbox{\gtsima}} 
\def\simlt{\lower.5ex\hbox{\ltsima}}

\def\msun{{\rm M}_{\odot}}
\def\lsun{{\rm L}_{\odot}}
\def\dsun{{\cal D}_{\odot}}
\def\fsun{\xi_{\odot}}
\def\zsun{{\rm Z}_{\odot}}
\def\msunyr{\msun {\rm yr}^{-1}}
\def\gdens{\msun\,{\rm kpc}^{-2}}
\def\sfrdens{\msun\,{\rm yr}^{-1}\,{\rm kpc}^{-2}}

\def\mum{\mu {\rm m}}
\newcommand{\angstrom}{\mbox{\normalfont\AA}}
\def\cc{\rm cm^{-3}}
\def\uflux{{\rm erg}\,{\rm s}^{-1} {\rm cm}^{-2} }

\def\fdust{\xi_{d}}
\def\fesc{f_{\rm esc}\,}
\def\td{\tau_{sd}}
\def\Sg{$\Sigma_{g}$}
\def\S*{$\Sigma_{\rm SFR}$}
\def\Ssfr{\Sigma_{\rm SFR}}
\def\Sgas{\Sigma_{\rm g}}
\def\Sstar{\Sigma_{\rm *}}
\def\Sesc{\Sigma_{\rm esc}}
\def\Srad{\Sigma_{\rm rad}}

\def\Dsolar{${\cal D}/\dsun$}
\def\Zsolar{$Z/\zsun$}
\def\DDsolar{\left( {{\cal D}\over \dsun} \right)}
\def\ZZsolar{\left( {Z \over \zsun} \right)}
\def\kms{{\rm km\,s}^{-1}\,}
\def\skms{$\sigma_{\rm kms}\,$}

\def\Scii{$\Sigma_{\rm [CII]}$}
\def\Sciimax{$\Sigma_{\rm [CII]}^{\rm max}$}
\def\CII{\hbox{[C~$\scriptstyle\rm II $]~}}
\def\CIII{\hbox{C~$\scriptstyle\rm III $]~}}
\def\OII{\hbox{[O~$\scriptstyle\rm II $]~}}
\def\OIII{\hbox{[O~$\scriptstyle\rm III $]~}}
\def\HH{\hbox{H$_2$}~} 
\def\HI{\hbox{H~$\scriptstyle\rm I\ $}} 
\def\HII{\hbox{H~$\scriptstyle\rm II\ $}} 
\def\CIion{\hbox{C~$\scriptstyle\rm I $~}}
\def\CIIion{\hbox{C~$\scriptstyle\rm II $~}}
\def\CIIIion{\hbox{C~$\scriptstyle\rm III $~}}
\def\CIVion{\hbox{C~$\scriptstyle\rm IV $~}}
\def\nhh{n_{\rm H2}}
\def\nhi{n_{\rm HI}}
\def\nhii{n_{\rm HII}}
\def\fhh{x_{\rm H2}}
\def\fhi{x_{\rm HI}}
\def\fhii{x_{\rm HII}}
\def\fd{f^*_{\rm diss}} 
\def\ks{\kappa_{\rm s}}

\def\cyan{\color{cyan}}
\definecolor{apcolor}{HTML}{b3003b}
\definecolor{afcolor}{HTML}{800080}
\definecolor{lvcolor}{HTML}{DF7401}
\definecolor{mdcolor}{HTML}{01abdf} 
\definecolor{cbcolor}{HTML}{ff0000}
\definecolor{sccolor}{HTML}{cc5500} 
\definecolor{sgcolor}{HTML}{00cc7a}

\title{Dusty winds clear JWST super-early galaxies}
\correspondingauthor{F. Fiore}
\email{fabrizio.fiore@inaf.it}

\author[0000-0002-4031-4157]{Fabrizio Fiore}
\affil{INAF-Osservatorio Astronomico di Trieste, via Tiepolo 11, 34143 Trieste, Italy}
\affil{IFPU - Institute for Fundamental Physics of the Universe, via Beirut 2, I-34151 Trieste, Italy}

\author[0000-0002-9400-7312]{Andrea Ferrara}
\affil{Scuola Normale Superiore, Piazza dei Cavalieri 7, 50126 Pisa, Italy}

\author[0000-0002-4314-021X]{Manuela, Bischetti}
\affil{Dipartimento di Fisica, Università di Trieste, Sezione di Astronomia, Via G.B. Tiepolo 11, I-34131 Trieste, Italy}

\author[0000-0002-4227-6035]{Chiara Feruglio}
\affil{INAF-Osservatorio Astronomico di Trieste, via Tiepolo 11, 34143 Trieste, Italy}

\author[0000-0002-8863-888X]{Andrea Travascio}
\affil{Universit\'a degli Studi di Milano Bicocca, Piazza dell’Ateneo Nuovo 1, 20126, Milan, Italy}

\begin{abstract}
The JWST discovery of a number of super-early (redshift $z>10$), blue galaxies requires these systems to be essentially dust-free in spite of their large stellar masses. A possible explanation is that dust is evacuated by radiatively-driven outflows. We test this hypothesis by deriving the Eddington ratio $\lambda_E=L_{\rm bol}/L_{E}$, where $L_{\rm bol}$ is the bolometric luminosity produced by star-formation and possible black hole accretion, for 134 galaxies at $6.5< z <16$. We find a strong anti-correlation between $\lambda_E$ and dust UV optical depth, $\tau_{1500} \propto \lambda_E^{-0.63}$; also, $\lambda_E$ increases with redshift. We confirm that galaxies exceeding a specific star formation rate ${\rm sSFR} > 13\, \rm Gyr^{-1}$ develop powerful outflows clearing the galaxy from its dust. This result is supported by ALMA dust continuum non-detections in three super-early systems. 
\end{abstract}

\section{Introduction} \label{sec:intro}
The first James Webb Space Telescope (JWST) images of the distant Universe are producing a tremendous progress in our understanding of the earliest phases of galaxy evolution. Such super-early (redshift $z \simgt 10$) galaxies show a large range of UV luminosities  ($L_{UV}=10^{42}-10^{46}$ ergs/s), stellar masses ($M_* \approx 10^{6-10}\rm M_\odot$), star-formation rates ($\rm SFR = 0.1-300\, \rm M_\odot \rm yr^{-1}$) and dust optical depths at rest-frame 1500\,\AA, $\tau_{1500}=0-15$ \citep{castellano2022,adams2023,furtak2022,topping2022,finkelstein2022,rodighiero2022,naidu2022,bradley2022,whitler2022,barrufet2022,trussler2022,leethochawalit2022,harikane2022,curti2022}. 

Remarkably, though, their sizes are quite similar and compact, with optical half-light radii, $r_e$, of just $0.1-0.5$ kpc \citep{yang2022,ono2022}. The corresponding high radiation energy density from bright star-forming sites, and possibly, accreting black holes (BHs), induces a strong radiation pressure on dust and gas. Depending on the Eddington ratio $\lambda_{E}\equiv L_{\rm bol}/L_{E}=$ (where $L_{\rm bol}$ is the bolometric luminosity produced by star-formation and any possible BH accretion, $L_E$ is the Eddington luminosity), and dust-to-gas ratio, $D$, radiation-driven dusty outflows might develop. Such outflows remove gas and dust from the galaxy, temporarily quenching its star formation until the next gas accretion event. Dusty outflows have been widely studied in the literature, using analytical, semi-empirical, and numerical approaches
\citep{ferrara1990,arav1994,scoville2001,thompson2005, murray2005,fabian2006, fabian2008, krum2012, krum2013,thompson2015, ricci2017}. The condition for the onset of the outflow can be expressed in terms of an \textit{effective} Eddington ratio for dust absorption, $\leff$, where the optical-UV dust extinction cross section, $\sigma_d$, substitutes the Thomson one, $\sigma_T$, in the standard $L_E$ definition. 

\begin{figure*}[t]
\label{distr}
\begin{tabular}{lcr}
\includegraphics[width=5.8cm]{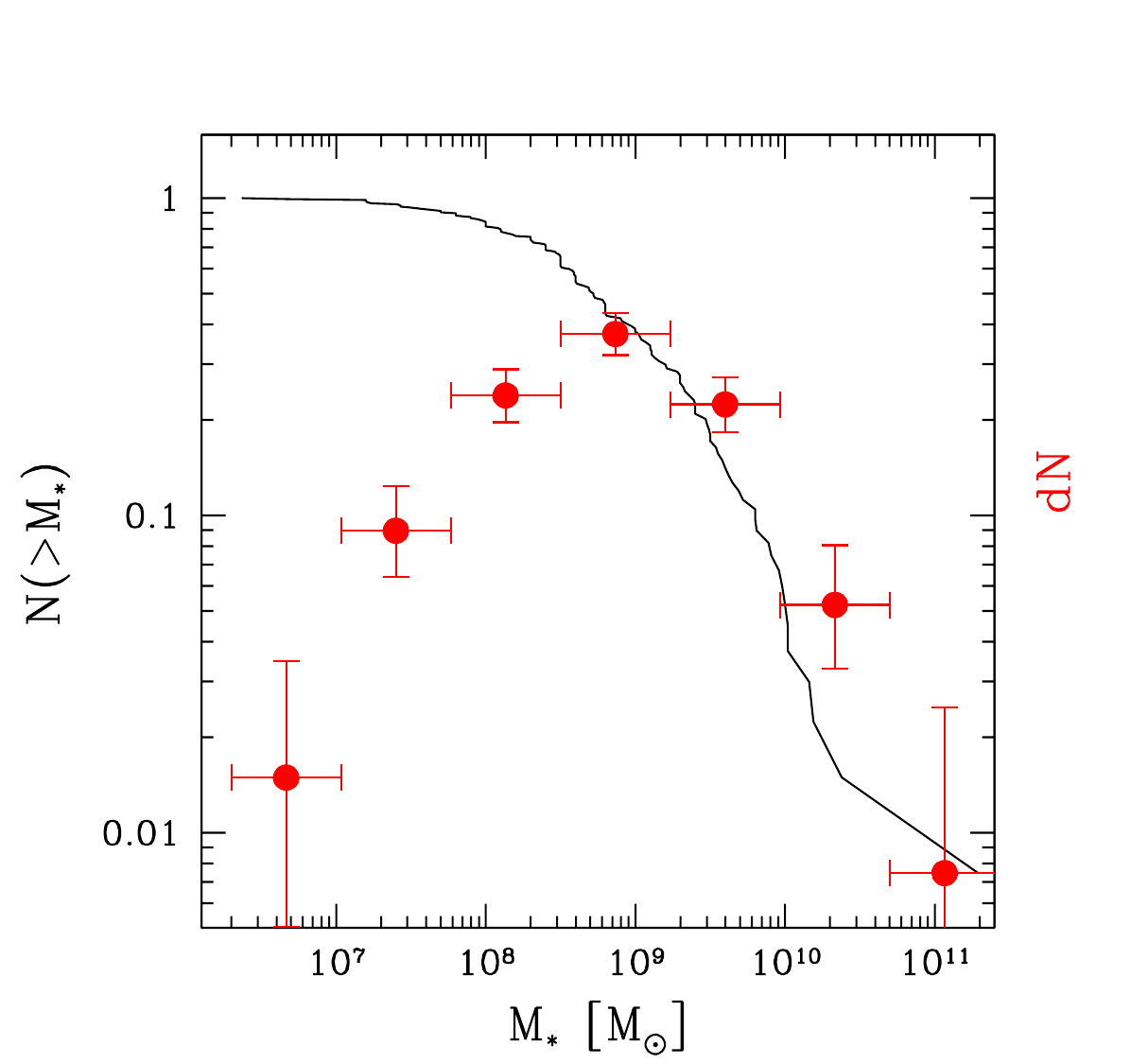}
\includegraphics[width=5.8cm]{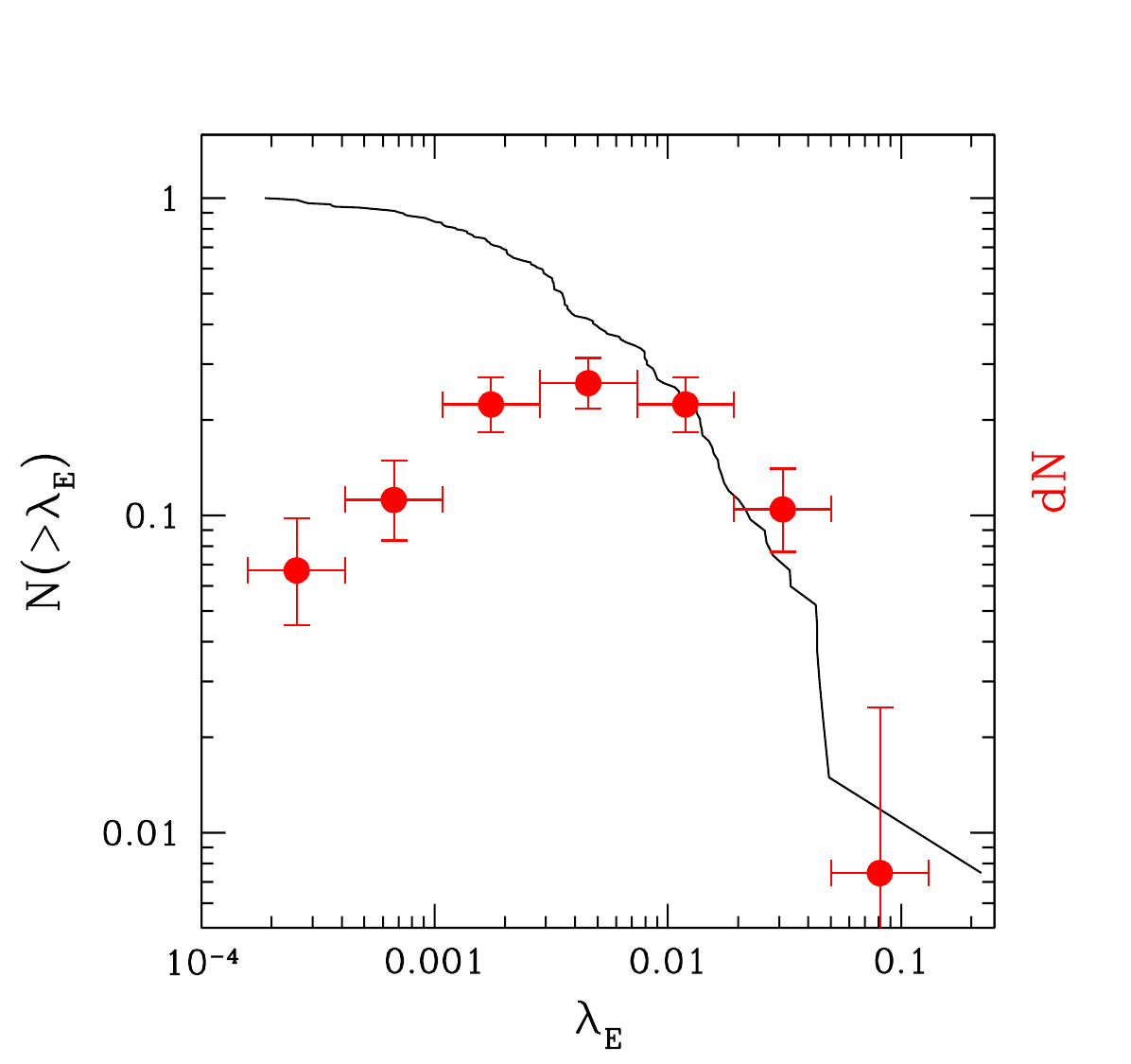}
\includegraphics[width=5.8cm]{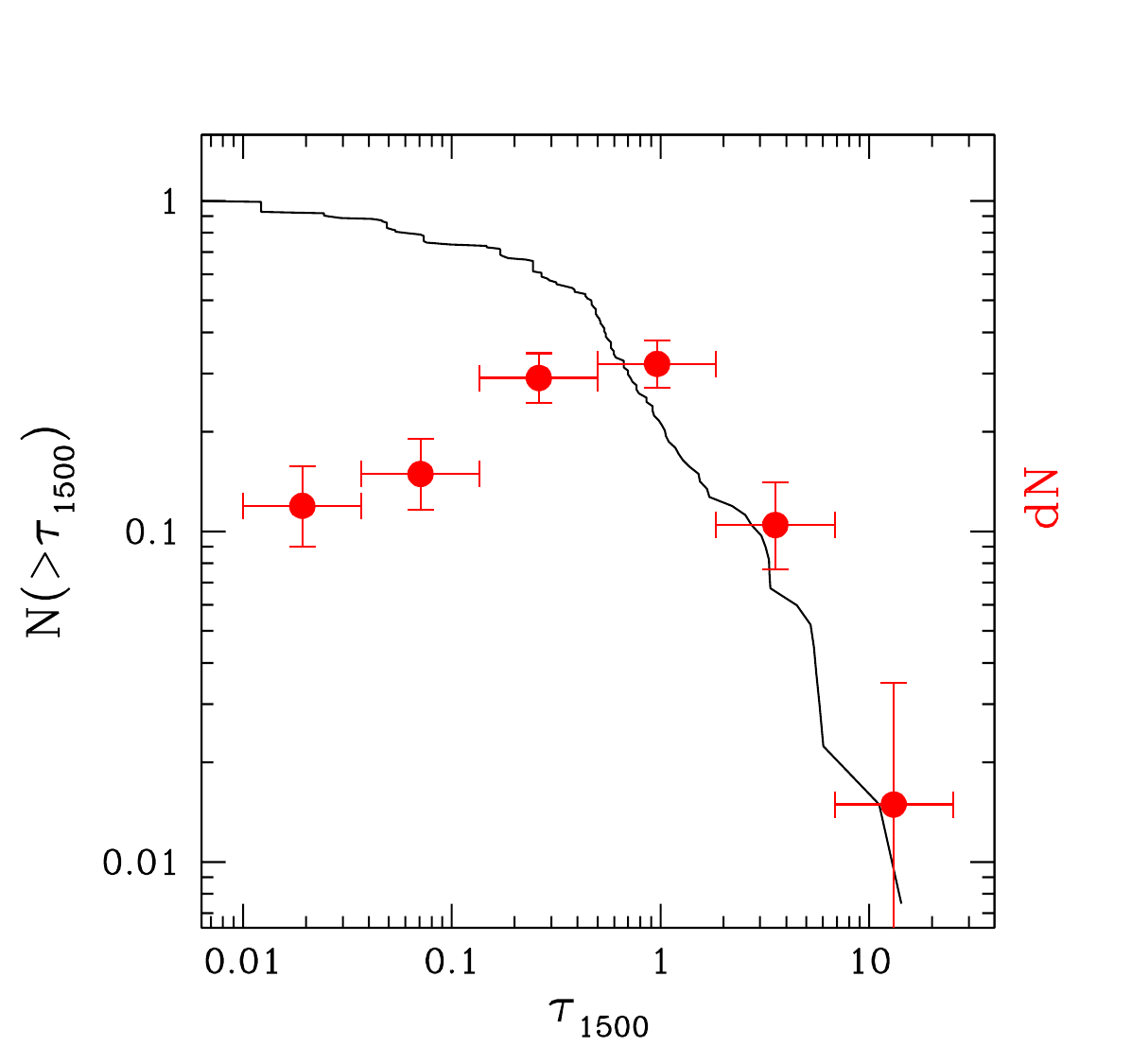}
\end{tabular}
\caption{Stellar mass, $M_*$, Eddington ratio, $\lambda_{E}$, and UV dust optical depth, $\tau_{1500}$, normalized cumulative (black curves) and differential (red points) distributions for our galaxy sample.}
\end{figure*}

If dust is considered, the radiation pressure efficiency is boosted by a factor $A=L_a/L_{\tau_T} \approx 10^{2-3}$ where $L_a$  is the absorbed luminosity for not fully ionized and dusty gas and $L_{\tau_T}$ is the absorbed luminosity for a fully ionized gas \citep[e.g.][]{fabian2006, fabian2008}. When $\lambda_E>\leff = \lambda_E/A$ radiation pressure is capable to expel dust and gas\footnote{We assume that dust and gas are strongly coupled by both Coulomb and viscous drag forces.} from the source. Dust ejection by radiation pressure has been indeed invoked by \citet{ziparo2022} to explain the blue colours of $z>10$ JWST candidates.

In this \textit{Letter} we collect a sample of 134 galaxies at redshift $6.5< z <16$, calculate their Eddington ratio, and compare it with $\leff$. We then interpret high-redshift galaxy observations in the framework of the radiation-driven dusty outflow scenario, expanding on \citet{ziparo2022} findings. 

\begin{figure*}
\begin{tabular}{cc}
\includegraphics[width=8.5cm,height=8.5cm]{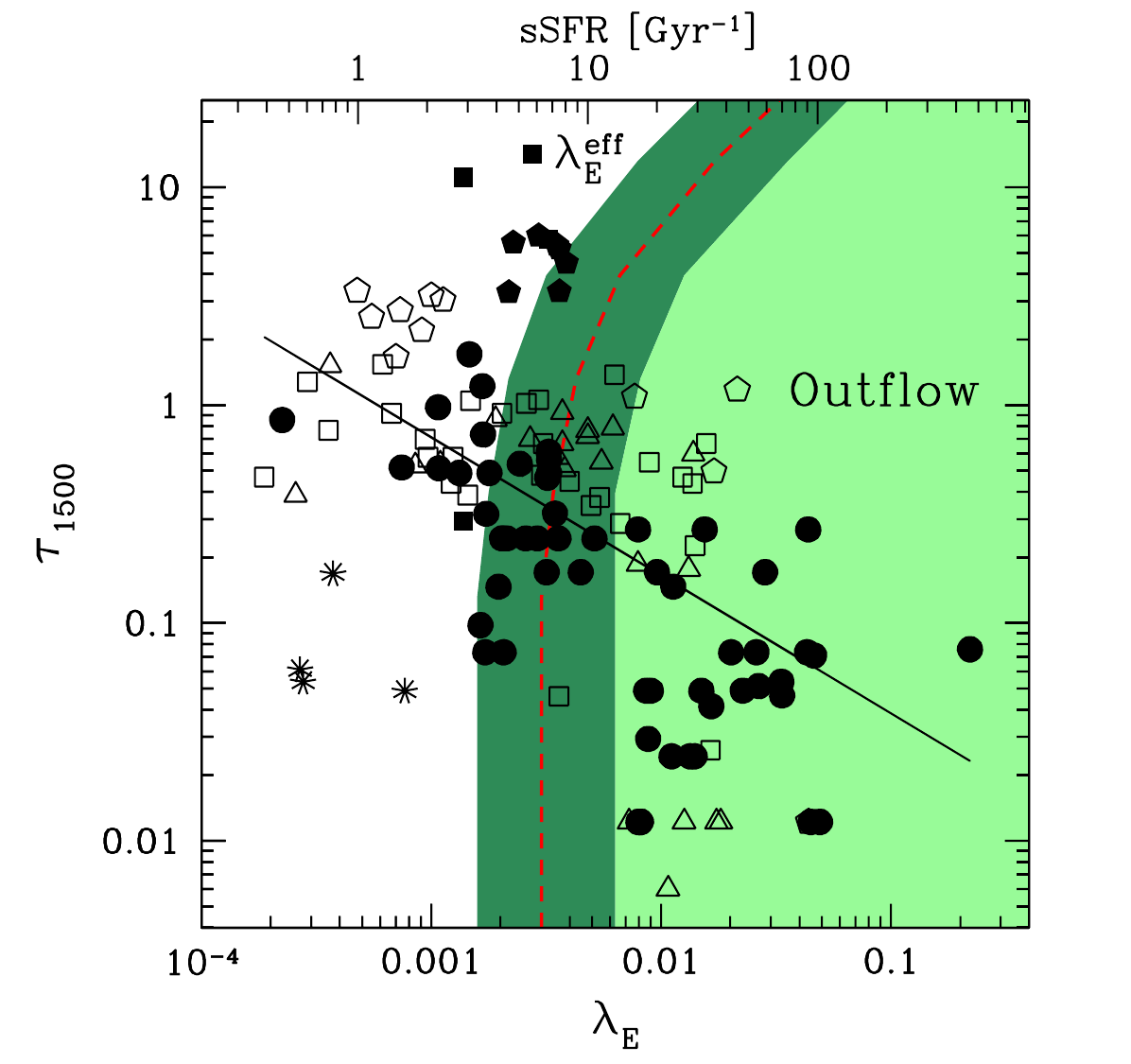}
\includegraphics[width=8.5cm,height=8.5cm]{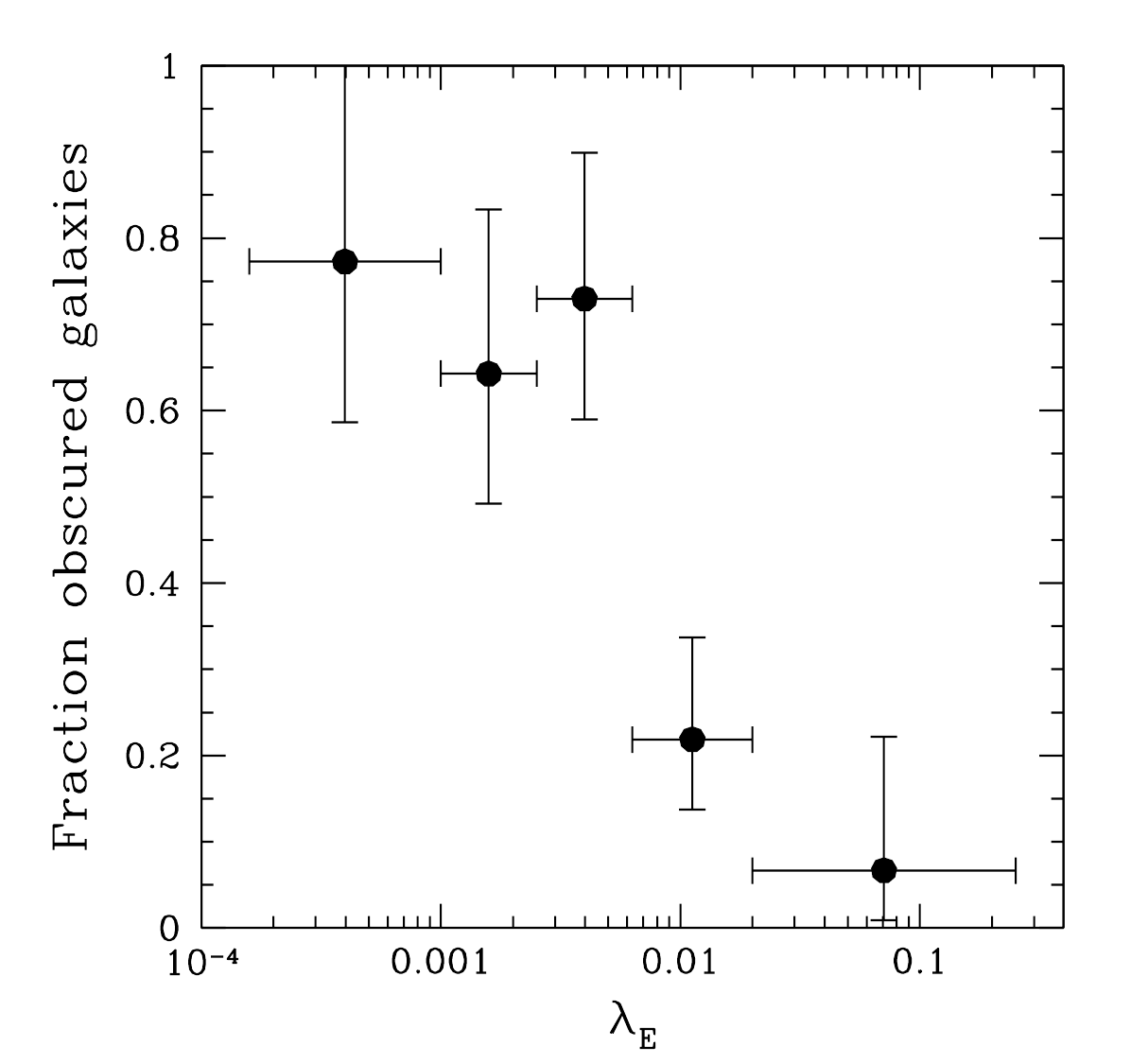}
\end{tabular}
\caption{\textit{Left panel}: Dust UV optical depth vs. Eddington ratio (lower axis) and specific SFR (upper axis) for 134 galaxies at $z>6.5$ in our sample. Filled circles are JWST-detected galaxy candidates from \citep{furtak2022,topping2022,finkelstein2022,naidu2022,bradley2022,whitler2022,trussler2022,leethochawalit2022,harikane2022,curti2022,castellano2022}. Filled squares identify candidates from \citet{rodighiero2022},  filled pentagons from \citet{barrufet2022}. Starred symbols are four sources from \citet{furtak2022} and \citet{bradley2022} for which the authors claim a Balmer break with ${\cal D}(4000)>2-3$. Open squares are REBELS sources (\citet{Bouwens2022}, open hexagons are REBELS sources with $\tau_{1500}$ estimated by \citet{ferrara2022}, open triangles are Ly$\alpha$ emitters from \citet{endsley2021}. The solid line is the best-fit power law $\tau_{1500} \propto \lambda_E^{-0.63}$; also shown is the adopted functional form for $\leff$ (red,dashed), the range of likely values (dark green), and the outflow region (light and dark green).  \textit{Right}: Fraction of galaxies with $\tau_{1500} > 0.4$ vs. Eddington ratio for our sample. }
\label{tauv}
\end{figure*}

\section{Sample selection}

We have collected from the recent literature extinction-corrected UV luminosities, SFR, stellar masses, and extinction properties (i.e., $\tau_{1500}$) for a sample of 134 galaxies at $z>6.5$. These quantities were computed in most cases by fitting the optical and near-IR photometry with a Spectral Energy Distribution (SED) model using the BEAGLE tool \citep{chevallard2016}. In particular, we have used data from \citet{furtak2022,topping2022,finkelstein2022,rodighiero2022,naidu2022,bradley2022,whitler2022,barrufet2022,trussler2022,leethochawalit2022,harikane2022,curti2022,castellano2022}. We complemented this JWST galaxy sample with the REBELS \citep{Bouwens2022} and CANDELS Ly$\alpha$ \citep{endsley2021} samples. 

We use published UV luminosities (at 1500\AA, $L_{1500}$) when available; in the other cases, these are obtained via the relation\footnote{ned.ipac.caltech.edu/level5/Sept12/Calzetti/Calzetti1\_2.html} $L_{1500}=2.2 \times 10^{43}\,{(\rm SFR/M_\odot \rm yr^{-1})}$ [erg s$^{-1}$], assuming a Kroupa stellar IMF with constant SFR over 100 Myr. We use  $\tau_{1500}$ when available, or obtain it from UV $\beta$ slopes as $\tau_{1500}=\beta-\beta_{\rm int}$ with $\beta_{\rm int}=-2.616$ \citep{reddy2018}, otherwise. From the definition $L_{E}=4\pi G m_p c M_*/\sigma_T = 1.26\times 10^{38}(M_*/M_\odot)$ergs/s, we compute the Eddington ratio as $\lambda_{E}=L_{\rm bol}/L_{E}$ by further assuming a bolometric correction, $f_{\rm bol} = L_{\rm bol}/L_{1500} = 2$, in agreement with the galaxy templates used in the next section to evaluate the boost factor. 

Fig. \ref{distr} shows the cumulative and differential distributions of $M_*,\ \lambda_{E}$ and $\tau_{1500}$. All of them span $3-4$ dex, a range much larger than possible statistical and systematic errors, thus describing intrinsic galaxy properties. The distributions follow a power-law at medium-high values, but show a turn-off below a certain threshold due to the sample incompleteness at low $M_*$. 

\section{Dusty outflows}
We evaluate the boost factor $A$ using \code{CLOUDY} \citep{2013RMxAA..49..137F} and \code{Starburst99} \citep{Leitherer99} simulations. We generate a large number of star-forming galaxy SEDs using Starburst99\footnote{https://www.stsci.edu/science/starburst99/docs/default.htm}, varying SFR, stellar mass, metallicity and age. We then use \code{Cloudy} to compute $A$, following \citet{fabian2006}. We found $A$ in the range $450-600$ for a $10-300$ Myr starburst with $\rm SFR = 1-100 M_\odot \rm yr^{-1}$, $M_*=10^{7-8}\, M_\odot$, metallicity $Z=0.6-3\, Z_\odot$, and a Galactic dust-to-gas ratio, $D=1/162$. $A$ reduces by $\sim40-45\%$ if a 30\% lower $D$ value is adopted. \citet{Arakawa2022} calculate $A$ for different dust grain composition and sizes finding values within the above ranges. At high optical depths two additional competing effects enters in the determination of $A$. On one side $A$ depends linearly on the optical depth of the absorbing gas because gas shells located beyond $\tau_{1500} \simgt 1$ are not subject to radiation pressure. On the other side in this regime reprocessing of the infrared radiation may further boost outflows \citep{Ishibashi2018}. The latter effect may be relevant for the galaxies with the highest densities ($N_H>10^{23}$cm$^{-2}$, $\tau_{1500}>50$), for which the infrared optical depth can exceed unity \citep{Pallottini2017}. 

Sources in the outflow regime should be virtually dust-free, and hence characterised by a small $A_V$, and blue UV spectral slopes ($\beta < -2$). On the contrary, systems that are not in that regime should be significantly attenuated by the dust produced by their stars. To test this basic hypothesis, in Fig. \ref{tauv} we plot $\tau_{1500}$ as a function of $\lambda_{E}$ or the specific SFR (left panel), and the fraction of sources with $\tau_{1500}>0.4$ (corresponding to 1 mag attenuation) as a function of $\lambda_{E}$ (right panel).

The optical depth $\tau_{1500}$ decreases with $\lambda_E$ following a power-law with index $-0.63\pm 0.10$; the Spearman Rank correlation coefficient is $R_{SR}=-0.53$ for 132 degrees of freedom, with a Student distribution $t=7.196$, implying a 100\% probability of correlation. The correlation is even more significant ($R_{SR}=-0.61$) if we exclude four sources (the starred symbols in Fig. \ref{tauv} left panel) for which \citet{furtak2022} and \citet{bradley2022}  claim a Balmer break with ${\cal D}(4000)>2-3$. Such high ${\cal D}(4000)$ value is unlikely at high-$z$ because the time to grow such a Balmer break would be higher than the Hubble time \citep[e.g.][]{maraston2005,noll2009}.
 We also note an abrupt change in the fraction of obscured sources when crossing the $\leff \approx 0.005$ line. Such fraction decreases from $80$\% for $\lambda_E \le \leff$ to $\approx 10$\% for larger values, where the outflow is predicted to occur.

\begin{figure*}
\begin{tabular}{cc}
\includegraphics[width=8.5cm,height=8.5cm]{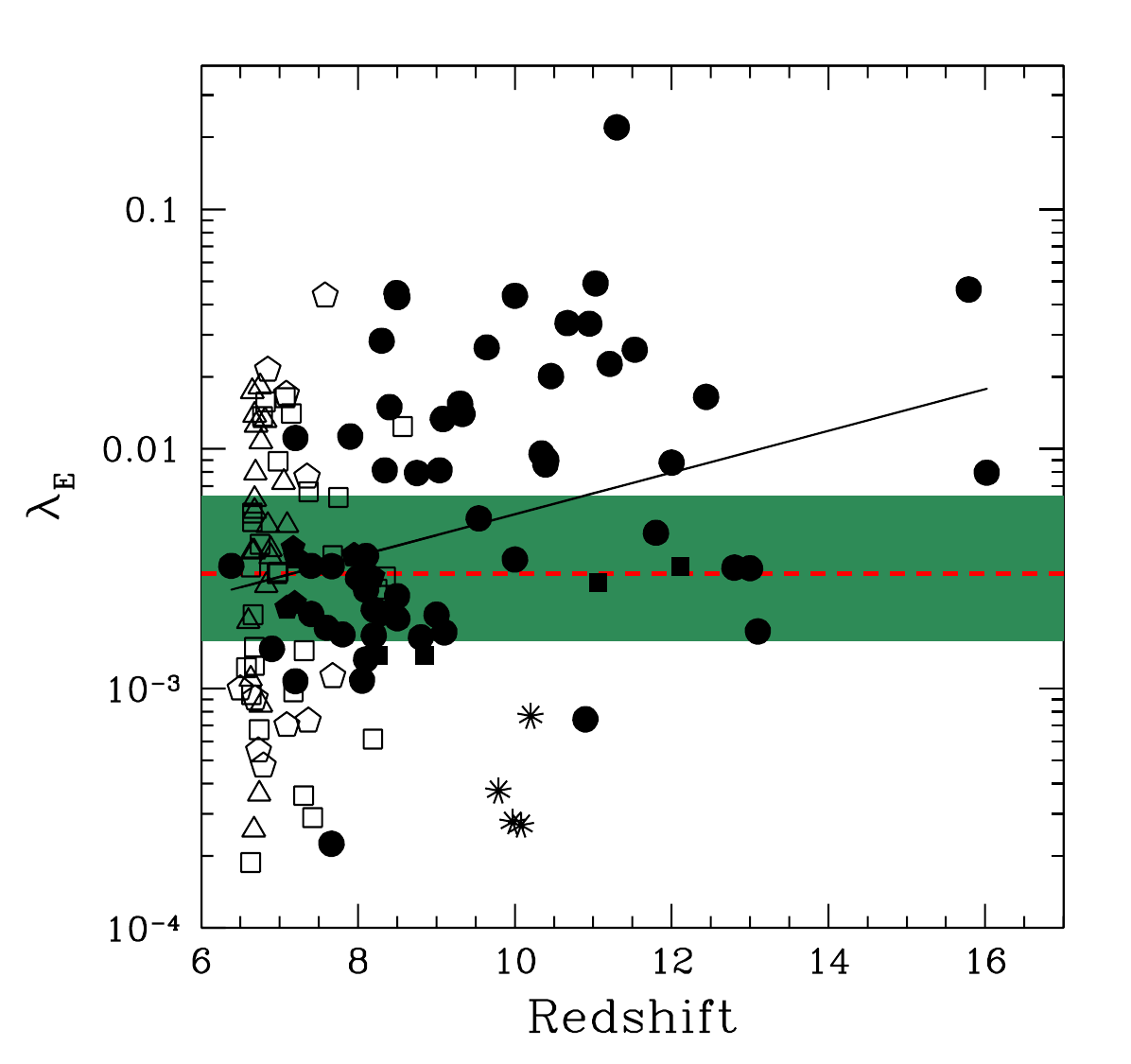}
\includegraphics[width=8.5cm,height=8.5cm]{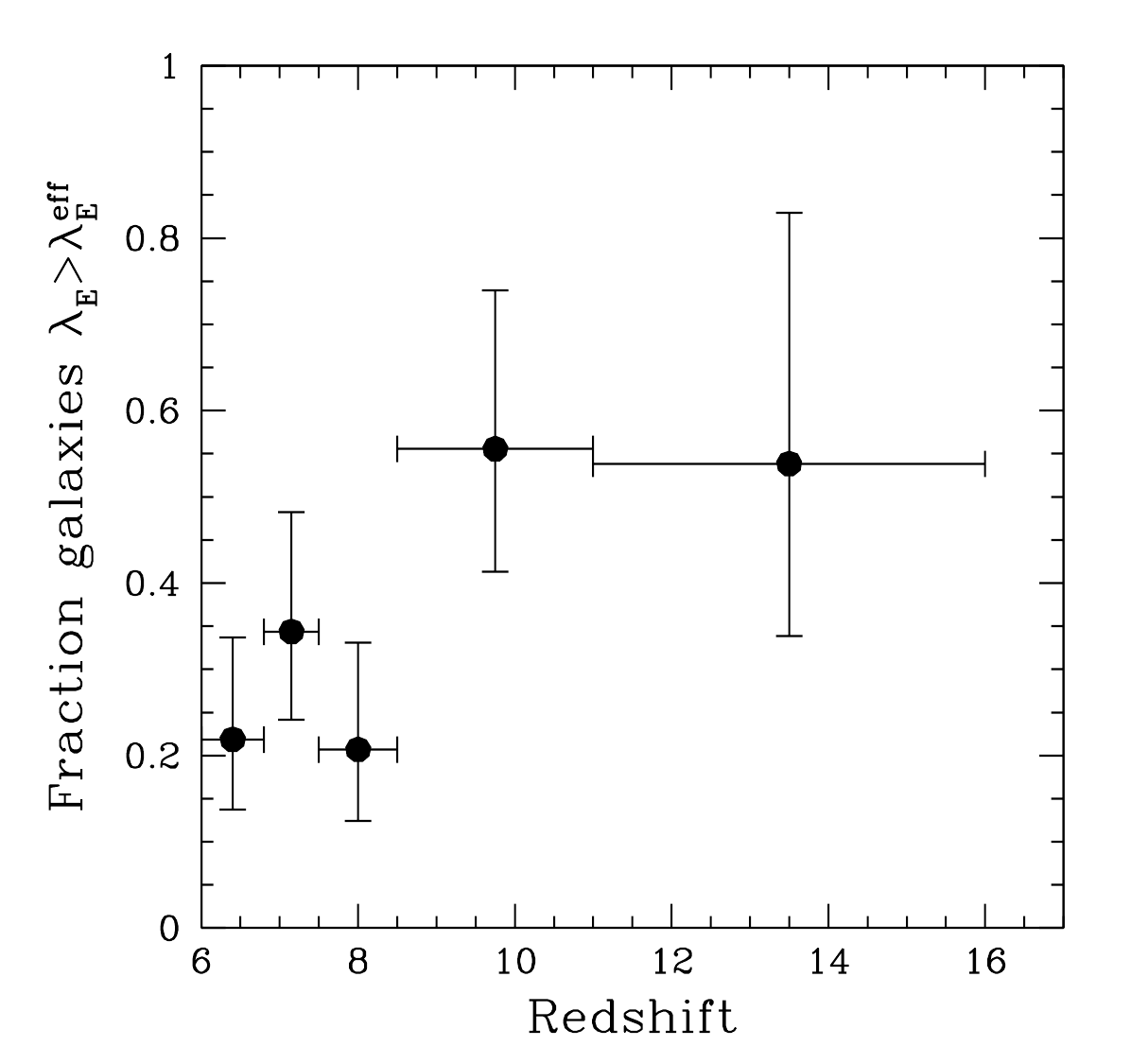}
\end{tabular}
\caption{\textit{Left panel}: Redshift evolution of the Eddington parameter for the galaxies in the sample. Points are the same as in Fig. 2. The solid line is the best-fit power-law, $\lambda_E \propto (1+z)^{2.1}$. Red dashed line marks the optically thin value of $\lambda_E^{eff}$, and the dark green region its most likely values.
\textit{Right}: Fraction of galaxies satisfying the outflow condition $\lambda_E > \leff$ vs. redshift.}
\label{lambdaz}
\end{figure*}

As a next step, we investigate the frequency of galaxies developing a dusty outflow with redshift (Fig. \ref{lambdaz}). From the left panel, we see that the Eddington ratio $\lambda_E$ increases with redshift, implying that early galaxies present conditions more favourable to the onset of radiation-driven outflows. The relation has a slope of $2.1\pm 0.6$, consistent with the \citep{Bouwens2022} determination. The Spearman Rank correlation coefficient of 0.242, and the probability of correlation is 99.74\%. 

Interestingly, the fraction of galaxies developing an outflow according to the condition $\lambda_E>\leff$, also increases with redshift, going from 20\% at $z<8.5$ to about 50\% at $z=8.5-16$. Thus, we conclude that a large fraction of the super-early galaxies detected by JWST are expected to have an outflow which has essentially emptied these systems of their interstellar medium.

\section{Discussion}
Despite early galaxies are complex systems, we have shown here that their observable properties are shaped by radiation-driven dusty outflows. Galaxies satisfying the outflow condition $\lambda_E>\leff$, are much less extincted than galaxies with lower $\lambda_E$. Moreover, the fraction of galaxies with $\lambda_E>\lambda_E^{eff}$, thus candidates for hosting powerful dusty outflows, increases from about 20\% at $z=6.5-8.5$ to about 50\% at $z=8.5-16$, confirming the sharp transition in galaxy properties proposed by \citet{ziparo2022} at $z\simeq 8$ to explain the blue colors of JWST detected super-early galaxies. Because the outflow occurs when the radiation pressure exceeds the gravitational pressure, the outflow velocity is greater than the escape velocity. Therefore, gas is inevitably lost from the galaxy in this scenario (even if it remained loosely bound, galaxies move, and thus gas would be stripped by ram pressure). This conclusion is supported by the recent ALMA dust continuum non-detections in three super-early galaxies \citep{Bakx22, Popping22, Yoon22, Kaasinen22}.  On the other hand, rejuvenation of the galaxy occur both if gas fall back into the galaxy, as in the supernova-driven shock model by \citep{Nath2022}, and by cosmological accretion of gas, restarting the SF. In the former case obscuration is a transition phase, preceding and following the blowout phase. In the latter case, the accretion rate for a $10^{11}M_\odot$ halo at z=10 is $\approx280 M_\odot$/yr. So in 50-100 Myr the galaxy should return to a cosmological Dark Matter / baryons ratio.

The physical interpretation of the above results is straightforward if we assume that the observed UV luminosity is produced by star formation. First, recall that $\lambda_E \propto L_{\rm bol}/L_E$. As, in addition, $L_{\rm bol} \propto \rm SFR$, and $L_E \propto M_*$, it turns out that $\lambda_E \propto {\rm SFR}/M_* \equiv {\rm sSFR}$. Hence, the key parameter deciding whether a galaxy develops an outflow is the specific star formation rate. The outflow condition $\lambda_E > \leff$ simply translates into the condition ${\rm sSFR} > {\rm sSFR}^* = 13\, \rm Gyr^{-1}$. It is interesting to note that the SFR densities of the high-z galaxies are between a few $M_\odot$/yr/kpc$^2$ end a few hundreds $M_\odot$/yr/kpc$^2$ (assuming spherical symmetry and radii of the order of 100pc), thus smaller than the Eddington-limited values in \citep{Perrotta2021, Diamond-Stanic2021} (a few thousands $M_\odot$/yr/kpc$^2$).

It is useful to compare this specific SFR threshold with the predictions of numerical simulations for early galaxies. \citet[see their Fig. 3]{pallottini2022} find that at $z \approx 8$, ${\rm sSFR} \simeq 100\,\gyr^{-1}$ for young ($t_\star \lsim 100\,\myr$), small  ($M_\star \lsim 10^8\msun$) galaxies, and ${\rm sSFR} \sim 10\,\gyr^{-1}$ for older ($t_\star \gsim 200\,\myr$) more massive ($M_\star \gsim 10^9\msun$) ones. Simulations by \citet{kannan2022} (see their Fig. 5) show that $\rm sSFR \approx 10\, \gyr^{-1}$ evolves very weakly from $z=8$ to $z=15$, and it is independent of stellar mass. We note that these simulations under-predict the abundance of luminous galaxies at $z>10$, so their sSFR might be underestimated. \citet[their Fig. 18]{behroozi2019} show that the average sSFR shows an increasing trend independently of the galaxy halo mass for $z\simgt 4$; interestingly, the curve crosses the sSFR$^*$ threshold at  $z\approx 8$. Finally, \citet{Bouwens2022}, by combining Hubble Ultra Deep Field and JWST NIRCam medium-band observations, find that $8 < z < 13$ galaxies have on average a high sSFR $\approx 24.5\, \rm Gyr^{-1}$. In spite of modelling and experimental uncertainties, it is clear that high-$z$ galaxies are located very close or above the sSFR$^*$ value required to drive an outflow. As a result, dusty outflows could be very common features in these systems. 

The power-law shape of the bright-end of the high-$z$ galaxy UV luminosity function \citep{bowler2020,donnan2022} suggests that BH accretion could substantially contribute to the UV luminosity of these systems. Indeed, at $z\sim 6$ the number density of galaxies in $-22 \le M_{\rm UV} \le -20$ \citep{bowler2020} is only a factor of a few higher than the AGN densities at the same magnitudes \citep{giallongo2019, orofino21}, suggesting a high AGN fraction at high-$z$. This can be hardly assessed using SED fitting techniques, although claims of AGN detection in JWST data using such techniques do exist \citep{onoue22}. Reliable estimates of AGN fractions will be soon provided by JWST spectroscopy via the detection of \NV and \HII lines in these systems. Furthermore, JWST spectroscopy could also provide observational evidences for outflows, allowing us to confirm and better quantify the scenario suggested in this letter. If most of the luminosity of the brightest high-$z$ galaxies is due to BH accretion, it requires BH masses $\approx10^7-10^8 M_\odot$ accreating at rates closer to their Eddington luminosities. In this case, galaxy cleaning is mostly due to BH accretion rather than star-formation.

Mechanical power can be provided to the outflow by supernova explosions. However, their contribution is likely to be sub-dominant as -- due to the expected very high gas density in these systems, the explosion energy is rapidly radiated away\footnote{For example, for the typical values of $M_*\approx 10^9 M_\odot$, $r_e = 100$ pc, and assuming a gas fraction $f_g = M_g/(M_g+M_*) \simgt 0.5$, appropriate for high-$z$ galaxies, we obtain a mean gas density $n \simgt 10^4 \cc$. In these conditions, free-free cooling would cause the supernova blast to become radiative after $\approx 100$yr, thus dramatically quenching its propagation \citep{Terlevich1992,Santiago2019}.} in a catastrophic cooling event \citep[][]{Pizzati20}. Similar conclusions concerning the inefficiency of supernova-driven outflows in early massive galaxies are presented in \citet[]{Bassini22}  and in the simulations of \citet{Nath2022}, which find that the shock timescale to cleanup a galaxy is 2-5 times longer that the SN dust production timescale. In the latter model gas ejected by supernova-driven shocks can fall back into the galaxy, giving rise to a new obscured phase.

\acknowledgments
We acknowledge support from PRIN MIUR 2017 Black hole winds and the baryon life cycle of galaxies, 2017PH3WAT. AF acknowledges support from the ERC Advanced Grant INTERSTELLAR H2020/740120. Generous support from the Carl Friedrich von Siemens-Forschungspreis der Alexander von Humboldt-Stiftung Research Award is kindly acknowledged (AF). 

%

\facilities{HST, JWST}


\software{astropy \citep{2013A&A...558A..33A},  
          CLOUDY \citep{2013RMxAA..49..137F}, 
          Starburst99 \citep{Leitherer99}
          }



\bibliography{bibliography} 




\end{document}